\documentclass[prd,10pt,a4paper,twocolumn,showpacs]{revtex4}
\usepackage{amsmath}
\usepackage{amssymb}
\usepackage{graphicx}
\usepackage{times}
\newcommand{\be}{\begin{equation}}
\newcommand{\ee}{\end{equation}}
\newcommand{\ben}{\begin{eqnarray}}
\newcommand{\een}{\end{eqnarray}}
\newcommand{\bes}{\begin{subequations}}
\newcommand{\ees}{\end{subequations}}

\newcommand{\bb}{\bibitem}

\begin{document}
\title{Lump-like Structures in Scalar-field Models}
\author{A.T. Avelar$^a$, D. Bazeia$^b$, W.B. Cardoso$^a$, and L. Losano$^b$}
\affiliation{{$^a$Instituto de F\'{\i}sica, Universidade Federal de Goi\'as, 74.001-970, Goi\^ania, Goi\'as, Brazil.}\\
{$^b$Departamento de F\'{\i}sica, Universidade Federal da Para\'\i ba, 58.059-900, Jo\~ao Pessoa, Para\'\i ba, Brazil.}}
\pacs{11.27.+d; 11.25.-w; 98.35.Gi; 98.80.Cq}

\begin{abstract}
In this work we investigate the presence of lump-like solutions in models described by a
single real scalar field. We take advantage of a procedure recently used to describe explicit
analytical solutions and we study several distinct models, showing how the parameters can be used
to control the specific features of the lump-like structures. The proposed models are of direct
interest to the construction of q-balls, to induce tachyonic excitations and gravitating structures of nontopological profile
on braneworld models with a single extra dimension, to map solitons in optical fibers, and to describe
collective excitations in Bose-Einstein condensates.
\end{abstract}

\maketitle


\section{Introduction}


In high energy physics, defect structures can be of topological or nontopological nature \cite{B1,B2}. Bell-shaped lump-like defects are nontopological structures which appear in models described by real scalar fields in $(1,1)$ spacetime dimensions under the action of nonlinear interactions. They are of interest in a diversity of situations, and in the present work we focus attention on the possibility of constructing lump-like structures with specific features controlled by small number of parameters, which appear in the potential to balance the nonlinearities that define the model. 

The presence of nonlinearities hardens the task of finding explicit analytical solutions. However, in this work we take advantage of the procedure set forward in \cite{AB}, which introduced a first-order framework that works very nicely to search for lump-like excitations. Possible applications are listed in \cite{AB}, and here we emphasize that the search for bell-shaped lump-like structures is of direct interest to the constructions of q-balls, which are stable structures due to the presence of bosonic charge, which enters the game through the presence of global symmetries of the enlarged model \cite{QB1}. 

Applications of current interest also include the use of lumps in the braneworld scenario with a single extra dimension of infinity extent. In this case, recent investigations deal with the presence of tachyonic states living on the brane \cite{FB}, the specific features of the gravitating nontopological structures in the five dimensional anti-de Sitter spacetime \cite{G}, and issues of stability and fermion modes living on such structures \cite{BD}. Several other applications are also of interest, and here we point out the direct use of lump-like models to describe the presence of bright solitons in optical fibers \cite{BS} and to mimic localized excitations in Bose-Einstein condensates \cite{ABC}.  

In this paper we introduce and study other new models described by a single real scalar field. We search for the lump-like solution, identifying the corresponding energy density, energy, amplitude and width as well. Before doing that, however, in the next Sec.~II we review some basic facts about the procedure introduced in \cite{AB}. We then use this in Sec.~III, where we investigate a new potential and also in Sec.~IV, where we study several distinct families of models, which present interesting solutions of the lump-like type. We end the work in Sec.~V with some comments
and conclusions.

\section{The framework}

We consider a single real scalar field in $(1,1)$ space-time dimensions, and use the Lagrange density
\begin{equation}
\mathcal{L}=\frac{1}{2}\partial _{\mu }\phi \,\partial ^{\mu }\phi -V(\phi ),
\label{sm}
\end{equation}%
where $V(\phi )$ is the potential, $x^{\mu }=(t,x)$ and $x_{\mu }=(t,-x).$ Here we rescale $\phi$, $x$ and $t$ in order to work with dimensionless fields and coordinates. We suppose that the potential engenders a set of critical points, $\{\bar{\phi}_{1},...,\bar{\phi%
}_{n}\},$ such that $(dV/d\phi)(\bar{\phi}_{i})=0$ and $V(\bar{\phi}_{i})=0$ for $i=1,2,...,n$. The equation of motion which follows from the above model
is given by
\be
\frac{\partial^2\phi}{\partial t^2}-\frac{\partial^2\phi}{\partial x^2}+\frac{dV}{d\phi}=0,
\ee
and for $\phi=\phi(x)$ being static configuration we get  
\begin{equation}
\frac{d^{2}\phi }{dx^{2}}=\frac{dV}{d\phi }.\label{em1}
\end{equation}
It can be integrated to give
\begin{equation}
\frac{d\phi }{dx}=\pm \sqrt{2V+c},  \label{em1phi}
\end{equation}%
where $c$ is a constant of integration, real.

We are searching for defect structures, and the energy density of the static solutions $\phi=\phi(x)$ is given by
\begin{equation}
\epsilon=\frac12\left(\frac{d\phi}{dx}\right)^2+V(\phi).
\end{equation}
It is the addition of two portions, known as the gradient and potential contributions, respectively. Thus, the static solutions have to obey the
boundary conditions
\begin{equation}
\lim_{x\to\pm\infty}\frac{d\phi}{dx}\to0
\end{equation}
in order to ensure finiteness of the gradient portion of the energy.

The presence of the set of critical points allows us to distinguish two kinds of defect structures: there may be topological or kink-like structures,
which in general connect two distinct but adjacent critical points, $\bar\phi_i$ and $\bar\phi_{i+1}$ with the boundary conditions
$\phi(x\to\pm\infty)\to\bar\phi_i$ and $\phi(x\to\mp\infty)\to\bar\phi_{i+1}$. 
But there may also be non-topological or lump-like structures, which in general requires a single critical point, with the boundary conditions
$\phi(x\to\pm\infty)\to\bar\phi_i$. These boundary conditions imply that $c$ in \eqref{em1phi} should vanish, as a necessary condition for the presence of finite energy solutions, leading to
\begin{equation}\label{em1phi0}
\frac{d\phi}{dx}=\pm\sqrt{2V}.
\end{equation}

These two first-order equations should be considered with care. The issue here is that the topological or kink-like
solutions are monotonic functions of $x$, so the derivative does not change sign and in \eqref{em1phi0} each sign identifies one equation. These two
equations lead to kink and anti-kink, respectively. However, the non-topological solutions are
trickier, since they are not monotonic. In fact, their first derivatives change sign at the arbitrary point $x_{0}$, which is the center of the
solution. Due to the translational invariance of the theory we may take $x_{0}=0,$ for simplicity. Thus, the derivative of the lump-like solutions
change sign at $x=0,$ and so we must understand the above equations \eqref{em1phi0} as suggested in Ref.~\cite{AB}:
\begin{equation}
\frac{d\phi }{dx}=\sqrt{2V}\;\mathrm{for}\;x>0\;\;{\&}\;\;\frac{d\phi }{dx}=-%
\sqrt{2V}\;\mathrm{for}\;x<0
\end{equation}%
and/or
\begin{equation}
\frac{d\phi }{dx}=-\sqrt{2V}\;\mathrm{for}\;x>0\;\;{\&}\;\;\frac{d\phi }{dx}=%
\sqrt{2V}\;\mathrm{for}\;x<0.
\end{equation}%
The presence of the and/or connection between the above pairs of equations can be understood with the help of the reflection symmetry: the two pairs of
equations appear when the model presents reflection symmetry; otherwise, we will only need a single pair of equations -- see below for further comments
on this issue.

The fact that the lump-like solutions are non-monotonic functions induces an important feature to the non-topological structures: the quantum mechanical problem engenders zero-mode with a node at the center of the structure, showing that there is at least one bound state with negative energy, indicating instability of the solutions. This fact hardens the study of non-topological or lump-like solutions in general. However, we can change this scenario, changing the way we investigate lump-like solutions. We first recognize that the lump-like structure is monotonic for $x$ positive, and for $x$ negative, separately.
Thus, for $x$ positive or for $x$ negative the non-topological profile is similar to the topological one, and so we can use symmetry arguments to
introduce $W=W(\phi)$ for non-topological solutions, with the understanding that the equations are now changed to
\begin{equation}  \label{EM11}
\frac{d\phi}{dx}=W_\phi\;\mathrm{for}\; x>0\;\;{\&}\;\;\frac{d\phi}{dx}%
=-W_\phi\;\mathrm{for}\; x<0
\end{equation}
and/or
\begin{equation}  \label{EM12}
\frac{d\phi}{dx}=-W_\phi\;\mathrm{for}\; x>0\;\;{\&}\;\;\frac{d\phi}{dx}
=W_\phi\;\mathrm{for}\; x<0.
\end{equation}
In this case, the energy associated with the lump-like solutions has the simple form
\begin{eqnarray}  \label{EL}
E_{\mathrm{l}}&\!\!=\!\!&|W(\phi(\infty))\!-\!W(\phi(0))|\!+\!|W(\phi(0))\!-
\!W(\phi(-\infty))|  \notag \\
&=&2|W(\phi(\infty))-W(\phi(0))|  \notag \\
&=&2|W(\phi(0))-W(\phi(-\infty))|.
\label{EW}
\end{eqnarray}

This is the first-order framework for non-topological or lump-like structures. It has two important advantages over the standard calculation: the
first is that it allows us to obtain the energy in a simple and direct manner, which does not depend of the explicit form of the solutions, as it is
required in the usual case. The second advantage is that one has to deal with first-order equations, instead of the second order equations of motion,
and this very much simplify the calculation. This fact motivates us to study new models, as we do in the next sections. For pedagogical reasons, however,
let us start with a simple model, before moving on the some family of models.

\begin{figure}[t]
\includegraphics[width=4cm]{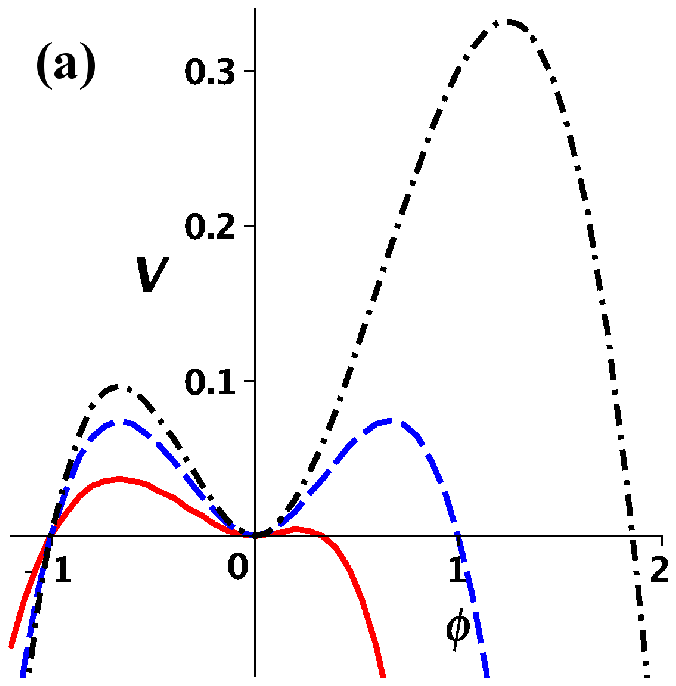}
\includegraphics[width=4cm]{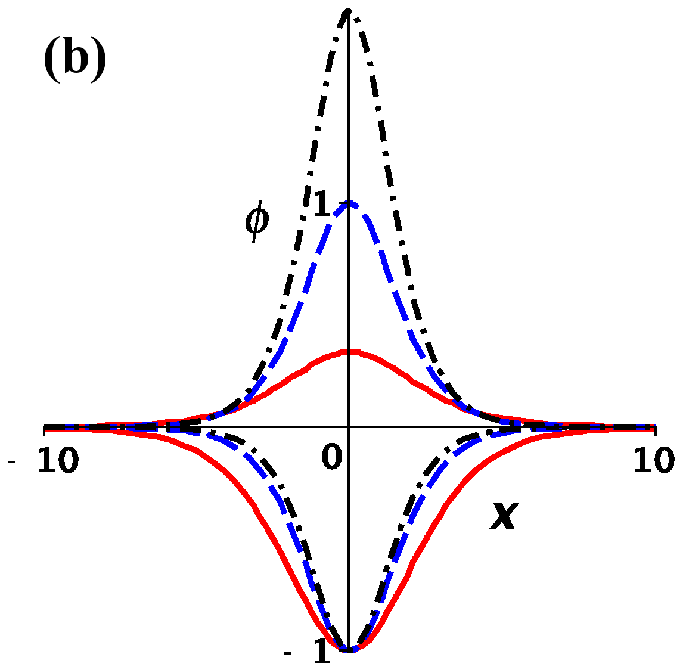}
\caption{(Color online) Plots of the (a) potential \eqref{p1} and the (b)
lump-like solutions \eqref{sp1-1} and \eqref{sp1-2}. The values of the parameter $a$ are $a=0.5$ for the solid (red) line,
$a=1$ for the dashed (blue) line, and $a=1.3$ for the dot-dashed (black) line.} \label{F1}
\end{figure}
   
\section{A simple model}

Firstly we consider a model described by the potential
\begin{equation}
V_{1}(\phi)=\frac{1}{2}\phi^{2}\left[a+(a-1)\phi-|\phi|\right],
\label{p1}
\end{equation}
where $a$ is positive parameter limited in the region $0\leq a<2$. This new model is described by the single parameter $a$, which
nicely controls the profile of the non-topological solutions. For $a=1$ it reduces to the symmetric inverted model introduced in
\cite{BI}, so it can be seen as a generalization of that model to the region $0\leq a<2$.

The above potential has three zeros, one at ${\bar\phi}=0$, which is the local minimum with $d^{2}V/d\phi^{2}=a$, and two others, at
${\bar\phi}_{1}=-1$ and at ${\bar\phi}_{a}=a/(2-a)$. In Fig. \ref{F1}a we plot the potential for some values of the parameter. The maxima of the
potential are located at ${\tilde\phi}_{1}=-2/3$ and at
\be
{\tilde\phi}_{a}=\frac{2a}{3(2-a)}.
\ee

To search for the solutions of the model described by the potential (\ref{p1}) we follow the procedure of the former Sec.~II. The presence of first-order equations very much simplify the calculations and we can write
\begin{equation}
\phi_{1}(x)=- \text{sech}^{2}\left( \frac{\sqrt{a}x}{2} \right),  \label{sp1-1}
\end{equation}
and
\begin{equation}
\phi_{a}(x)=\frac{a}{2-a} \text{sech}^{2}\left(\frac{\sqrt{a}x}{2} \right).
\label{sp1-2}
\end{equation}

\begin{figure}[t]
\includegraphics[width=4cm]{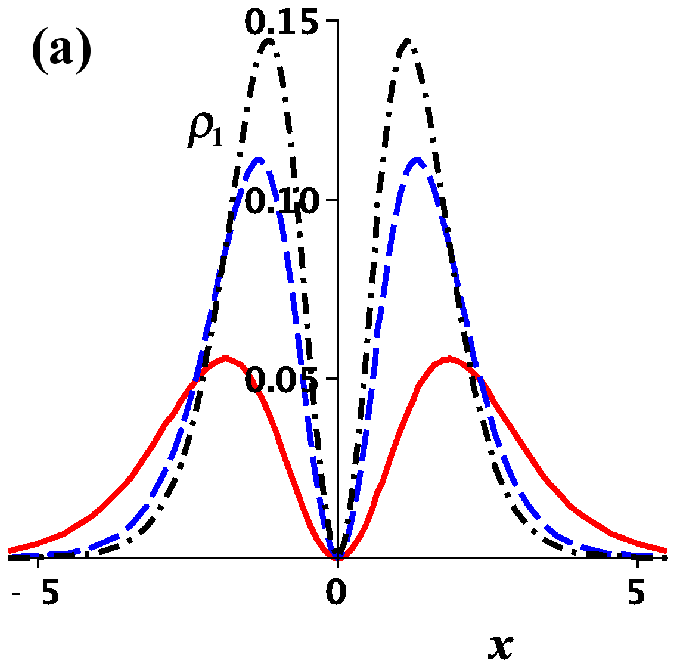}
\includegraphics[width=4cm]{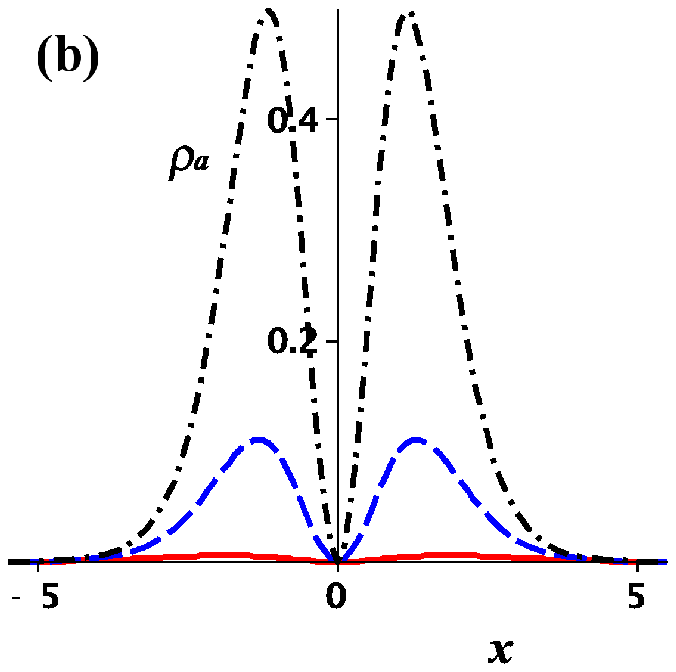}
\newline
\includegraphics[width=4cm]{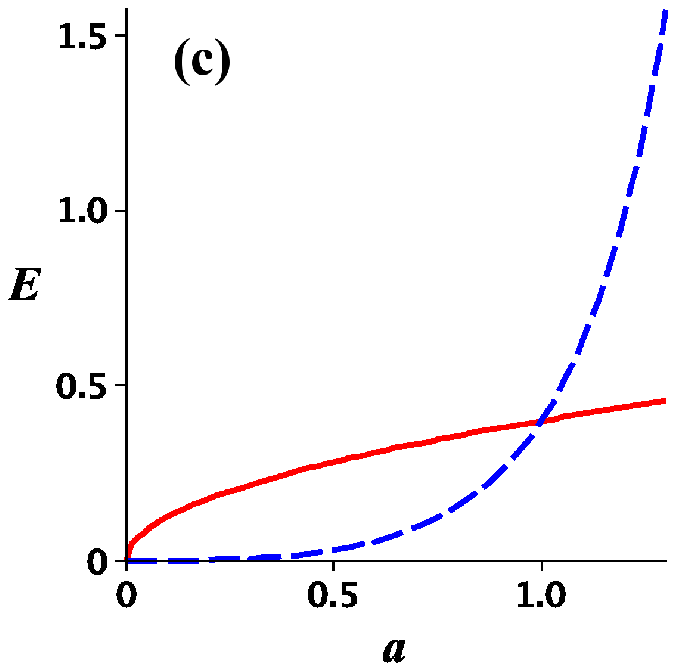}
\caption{(Color online) Plots of the energy densities of the lump-like solutions \eqref{sp1-1} in (a) and \eqref{sp1-2} in (b), for the same values of the parameter $a$ given in Fig.~1. In (c) we illustrate how the total energies $E_{1}$ and $E_{a}$ [see \eqref{Eaa}] vary with the parameter $a$, with solid (red) line and dashed (blue) line, respectively.}\label{F2}
\end{figure}

The subscripts $1$ and $a$ are related with the two sectors of the potential. They reduce to the lumps $\phi(x)=\pm {\rm sech}^{2}(x/2)$, for $a=1$. The lump $\phi_{1}(x)$ has its minimum at $\phi_{1}(0)=-1$, and $\phi_a(x)$ at $\phi_{a}(0)=a/(2-a)$. Both lumps have asymptotic values $\phi_{1}(\pm \infty)=\phi_{a}(\pm \infty)=0$. In Fig. \ref{F1}b, we show both solutions, and there we see that they have standard bell-shape profiles, but the $\phi_{1}$ solutions have constant amplitude and varying width, and the $\phi_{a}$ solutions have varying amplitude and width. Their energy densities are given by
\begin{equation}
\rho_{1}=a \text{sech}^{4} \left(\frac{\sqrt{a}x}{2} \right) {\tanh}^{2}
\left(\frac{\sqrt{a}x}{2} \right),
\end{equation}
\begin{equation}
\rho_{a}=\frac{a^{3}}{(a-2)^{2}} {\rm sech}^{4}\left(\frac{x\sqrt{a}}{2}%
\right){\tanh}^{2}\left(\frac{x\sqrt{a}}{2}\right),
\end{equation}
which are plotted in Fig.~\ref{F2}a and \ref{F2}b, respectively. The energies $E_{1}$ and $E_{a}$ corresponding to the above densities are given by
\be
E_{1}=\frac{8}{15}\sqrt{a},\;\;\;\;\;E_{a}=\frac{8}{15}(2-a)^{2}.
\label{Eaa}
\ee
We can check that the limit $a\rightarrow1$ sends both $E_{1}$ and $E_{a}$ to the same value $8/15$, as is shown in Fig. \ref{F2}c. The
above lump-like solutions have amplitude and width given by $A_{1}=1$ and
\be
L_{1}=\frac{4}{\sqrt{a}}\,{\rm arcsech}(1/\sqrt{2}),
\ee
and
\be
A_{a}=\frac{a}{2-a},\;\;\;\;\; L_{a}=\frac{4}{\sqrt{a}}\,{\rm arcsech}(1/\sqrt{2}).
\ee
Here we note that the limit $a\rightarrow1$ changes $A_{a}\rightarrow A_{1}$. Note also that $L_{a}=L_{1}$ for all values of $a$. This is interesting since for the solutions $\phi_a$, the parameter $a$ controls amplitude and energy, without changing the corresponding width. These lump-like solutions are interesting and may be useful in applications in several branches of nonlinear science. 

\section{Other models}

Let us now investigate other models, which represent some new families of models. Below we introduce three new families,
each one having specific features which make the lump-like solutions interesting for applications in high energy physics
and in condensed matter.

\subsection{First family}

\begin{figure}[t]
\includegraphics[width=4cm]{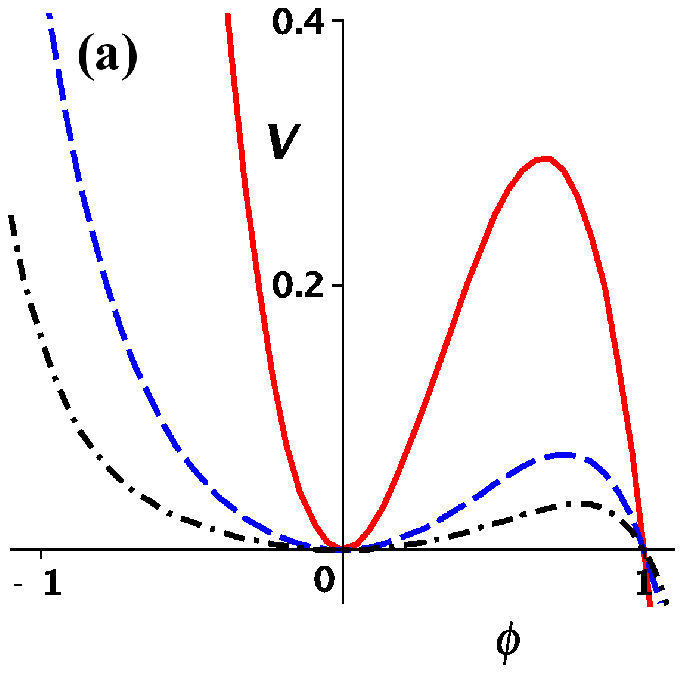}
\includegraphics[width=4cm]{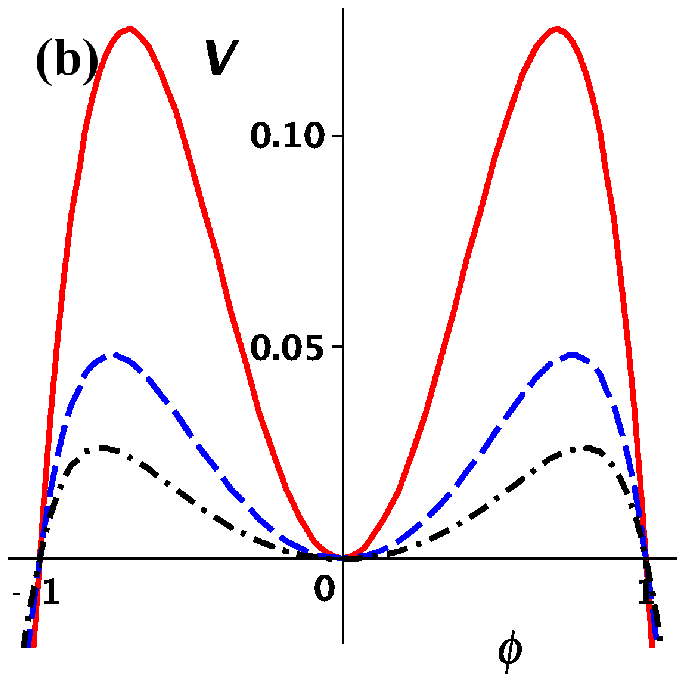}
\caption{(Color online) Plots of the potential (\ref{pqn}) for (a) $n$ odd and for (b) $n$ even. For $n$ odd we use $n=1$ for the solid (red) line, $n=3$ for the dashed (blue) line, and n=5 for the dot-dashed (black) line. For $n$ even we use $n=2$ for the solid (red) line, $n=4$ for the dashed (blue) line, and $n=6$ for the dot-dashed (black) line.}
\label{F3}
\end{figure}

Here we introduce the first family of models, which is described by the potential
\begin{equation}
V_{2}(\phi)=\frac{2}{n^{2}}\phi^{2}\left( 1-\phi^{n}\right),
\label{pqn}
\end{equation}
where we are assuming that $n$ is some positive integer. Note that for $n=1$ this potential is the $\phi^{3}$ model, and for $n=2$ it is
the inverted $\phi^{4}$ model \cite{BLM}. For other values of $n$ we get new models, and the potentials have distinct behaviors, one for $n$ odd and another one for $n$ even, as we illustrate in Fig~\ref{F3}. In general, the local maxima of the potentials are located at
\be
{\tilde\phi}_n=k\left(\frac{2}{2+n}\right)^{1/n},
\ee
where $k=1$ for $n$ odd and $k=\pm1$ for $n$ even. 

This potential presents two zeros for $n$ odd (at ${\bar\phi}=0,1$) and three for $n$ even (at ${\bar\phi}=0,\pm1$), as shown in Fig.~\ref{F3}. The minimum at $\phi=0$ leads to $d^2V/d\phi^2(\phi=0)=4/n^2$, which represents the classical mass (squared) of the bosonic excitation around the minimum $\phi=0$.

\begin{figure}[tbp]
\includegraphics[width=4cm]{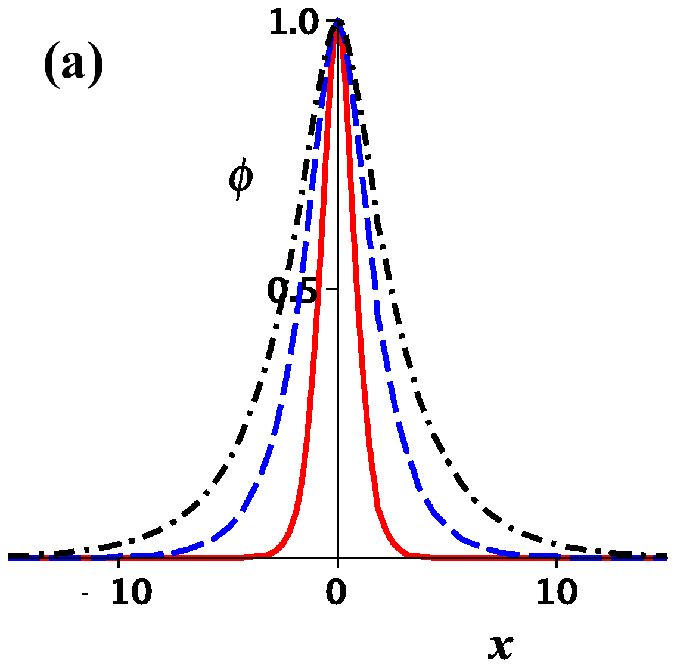}
\includegraphics[width=4cm]{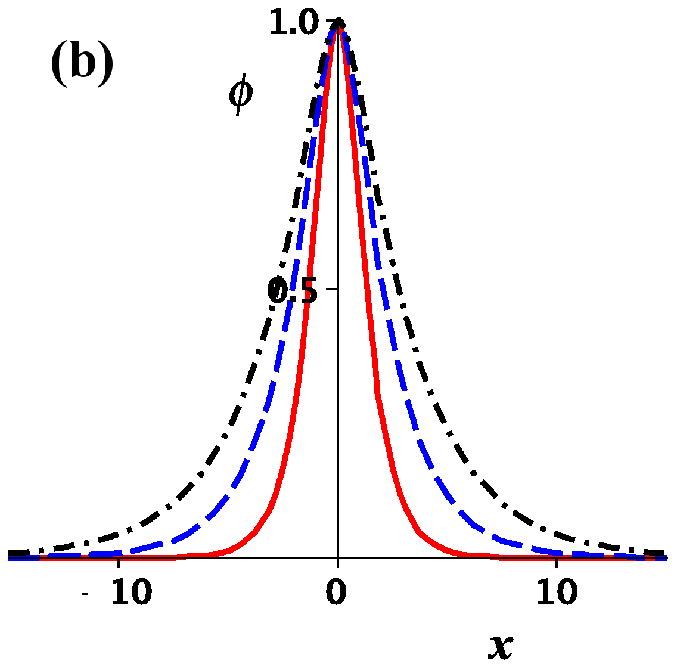}
\caption{(Color online) Lump-like solutions given by Eq.(\ref{sol}) for (a) $n$ odd and for (b) $n$ even. We consider the same conventions used in Fig.~3.}
\label{F4}
\end{figure}

\begin{figure}[tbp]
\includegraphics[width=4cm]{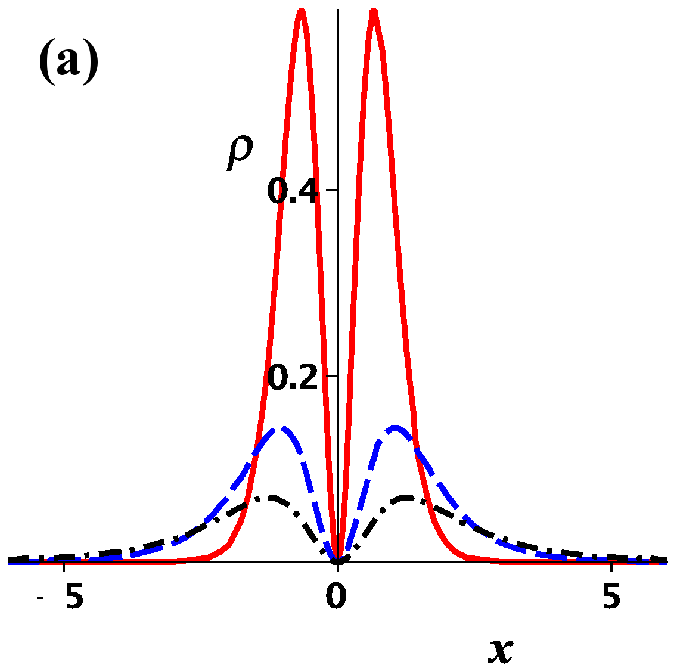}
\includegraphics[width=4cm]{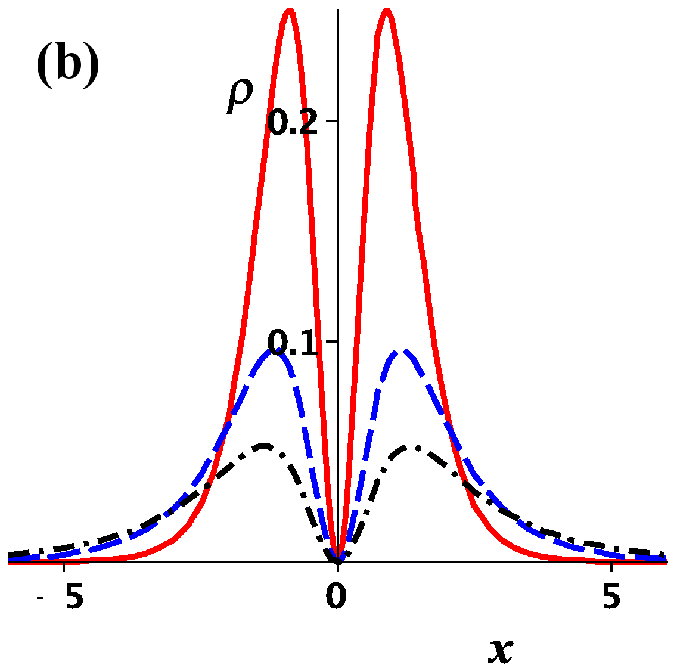}
\newline
\includegraphics[width=4cm]{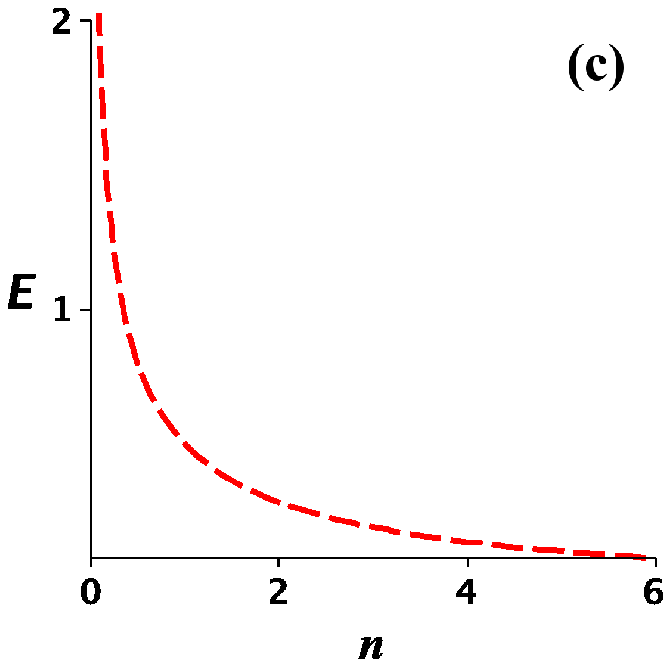}
\caption{(Color online) Energy density given by Eq.~\eqref{sol1} for (a) $n$ odd and for (b) $n$ even. We consider the same conventions used in Fig.3.
In (c) we display the total energy given by Eq.(\ref{Eb}) versus $n$.}\label{F5}
\end{figure}

To search for solutions we follow the procedure of Sec.~II. The family of models described by the potential \eqref{pqn} has lump-like solutions and energy densities which can be represented by
\begin{equation}
\phi(x)={\rm sech}^{2/n}(x),
\label{sol}
\end{equation}
and
\begin{equation}
\rho(x)=\frac{4}{n^{2}}{\rm sech}^{4/n}(x){\tanh}^{2}(x),
\label{sol1}
\end{equation}
respectively. The amplitude and width of the non-topological bell-shaped solution are $A_{n}=1$ and 
\be
L_{n}=2{\rm arcsech}\left(2^{-n/2}\right).
\ee
The solutions for some values of $n$ odd and even are shown in Figs.~\ref{F4}a and \ref{F4}b, respectively. Also, in the Figs.~\ref{F5}a and \ref{F5}b we plot their energy densities for some values of $n$, odd and even, respectively.

The total energy is derived from Eq.~\eqref{EW}; it has the form 
\begin{equation}
E(n)=\frac{\sqrt{\pi}}{2n}\frac{\Gamma\left(\frac{n+2}{n}\right)}{\Gamma\left(\frac{3n+4}{2n}\right)},
\label{Eb}
\end{equation}
where $\Gamma $ stands for the gamma function. In Fig.~\eqref{F5}c we plot the total energy as a function of $n$.

\subsection{Second family}

\begin{figure}[t]
\includegraphics[width=4cm]{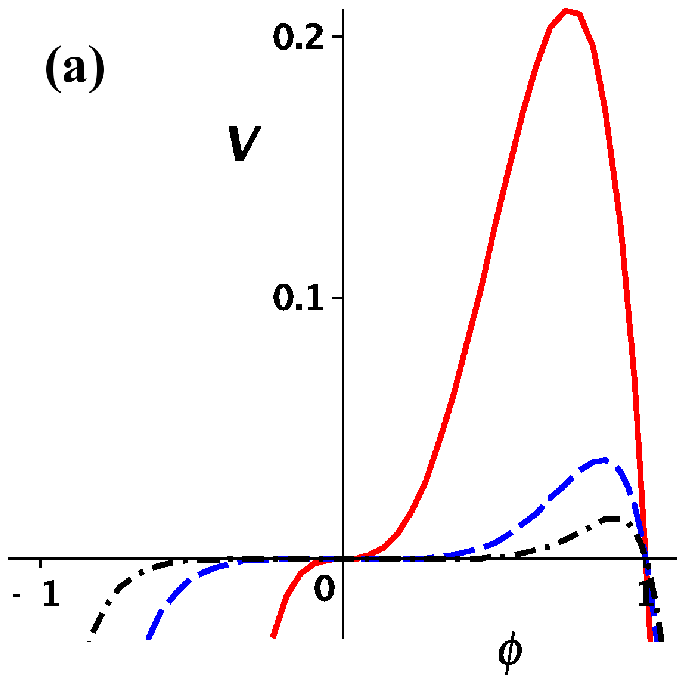}
\includegraphics[width=4cm]{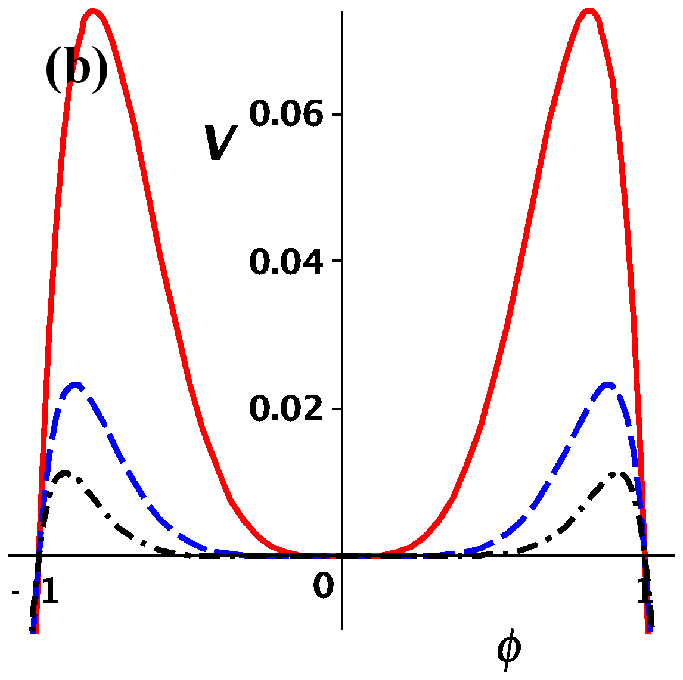}
\caption{(Color online) Plots of the potential (\ref{V3}) for (a) $n$ odd and for (b) $n$ even, depicted with the same conventions used in Fig.~3.}
\label{F6}
\end{figure}

Let us now introduce another family of models, described by the potential
\begin{equation}
V_{3}(\phi)=\frac{2}{n^2}\phi^{2+n}\left(1-\phi^{n}\right),
\label{V3}
\end{equation}
where $n$ is positive integer. The local maxima are at 
\be
{\tilde\phi}_n=k\left(\frac{n+2}{2n+2}\right)^{1/n},
\ee
with $k=1$ for $n$ odd and $k=\pm1$ for $n$ even. In Fig.~\ref{F6} we display the
potential (\ref{V3}) for $n$ odd and even, respectively.

This potential presents two zeros for $n$ odd (at ${\bar\phi}=0,1$) and three for $n$ even (at ${\bar\phi}=0,\pm1$) according to Fig.~\ref{F6}. We also have $d^2V/d\phi^2(\phi=0)=0$.

\begin{figure}[tbp]
\includegraphics[width=4cm]{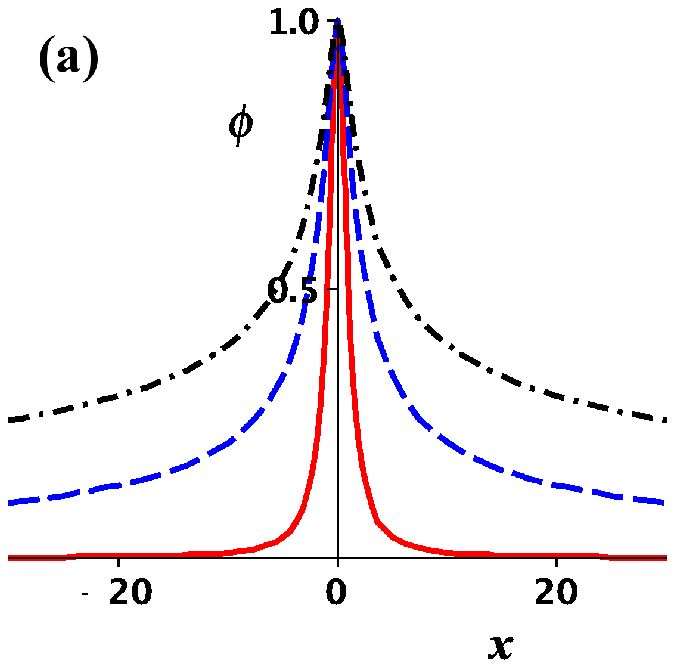}
\includegraphics[width=4cm]{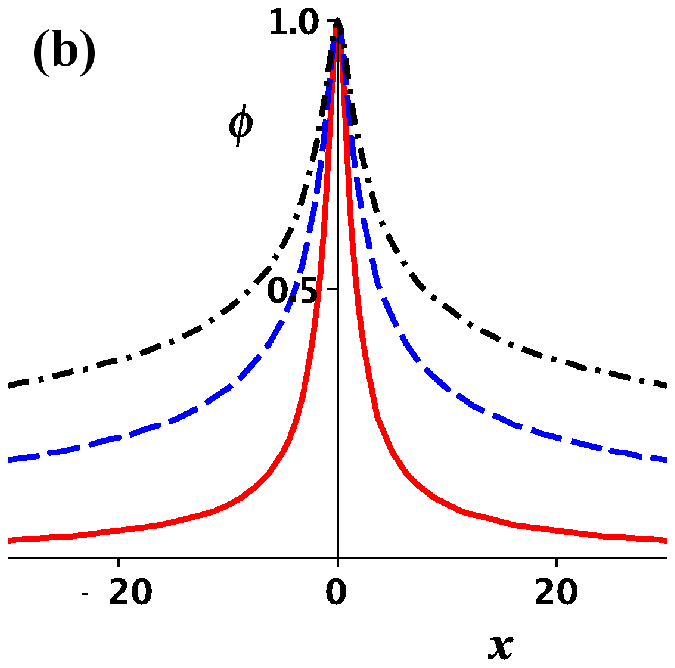}
\caption{(Color online) Plots of the lump-like solutions given by Eq.~\eqref{solV3} for (a) $n$ odd and for (b) $n$ even. We consider the same conventions used in Fig.~3.}\label{F7}
\end{figure}

\begin{figure}[tbp]
\includegraphics[width=4cm]{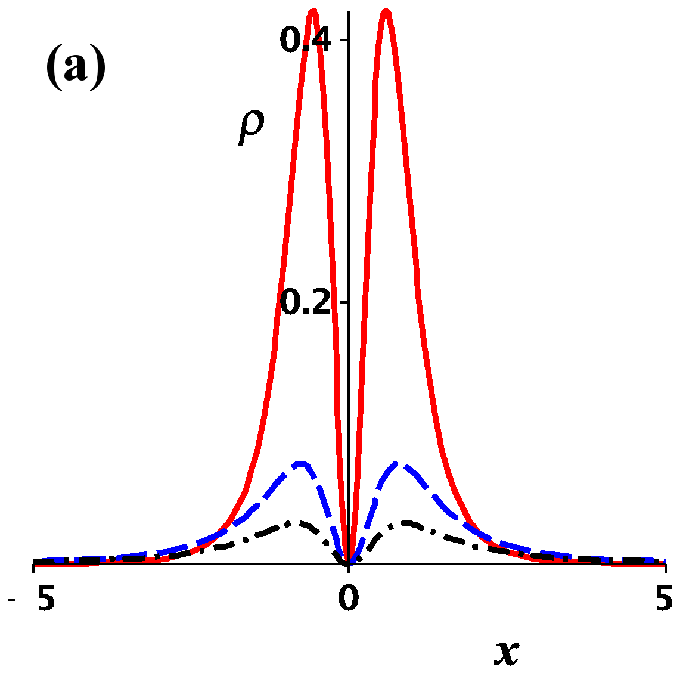} 
\includegraphics[width=4cm]{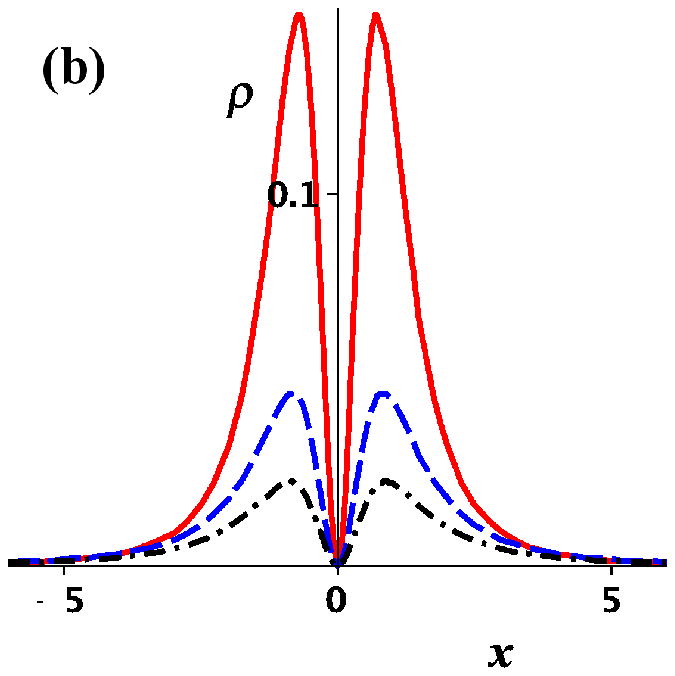}
\newline
\includegraphics[width=4cm]{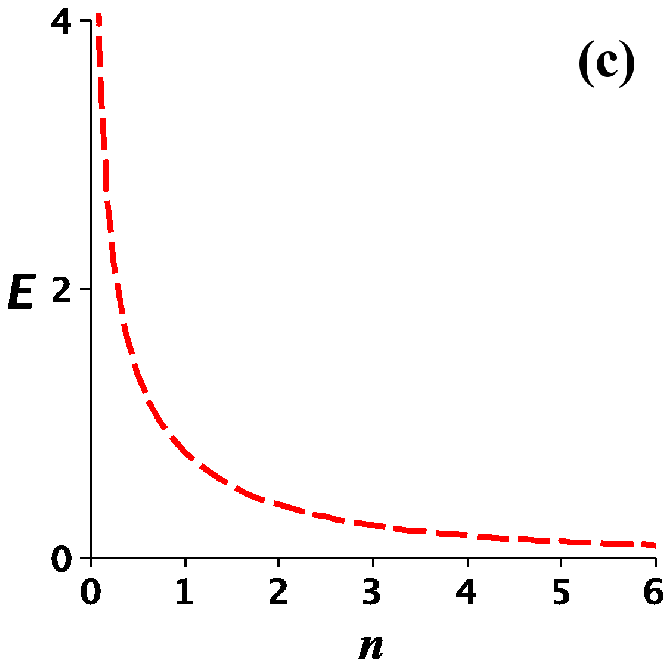}
\caption{(Color online) Energy density given by Eq.~\eqref{rhoV3} for (a) $n$ odd and for (b) $n$ even. We consider the same conventions used in Fig.~3.
In (c) we display the total energy given by Eq.~\eqref{EV3} versus $n$.} \label{F8}
\end{figure}

To search for solutions we proceed as before. The family of models described by \eqref{V3} has lump-like solution and energy density given by
\be
\phi(x)=\frac1{(1+x^{2})^{1/n}}.
\label{solV3}
\ee
and
\ben
\rho(x)=\frac{2}{{n}^{2}}\left[\left(1\!+\!{x}^{2}\right)^{-{\frac{2+2n}{n}}}\!({x}^{2}\!-\!1)\!+\!\left(1\!+\!{x}^{2}\right)^{-{\frac{n+2}{n}}}\right].\;
\label{rhoV3}
\een
The amplitude and width of the non-topological bell-shaped solution are $A_{n}=1$ and 
\be
L_{n}=\sqrt {{2}^{n}-1}.
\ee 
The solutions for $n$ odd and even are presented in Figs.~\ref{F7}a and \ref{F7}b, respectively. In Figs.~\ref{F8}a and \ref{F8}b we show their energy densities. We note that in this case the lump-like solutions are more diffuse, since they go to zero in a much slower way, if compared to the former case, for the first family of models.

The total energies of these lump-like solutions are given by
\begin{equation}
E(n)=\frac{\sqrt{\pi}}{\left(n+2\right)}\frac{\Gamma\left(1/2\,{\frac{n+4}{n}}\right) }{\Gamma\left(2\,/{n}\right)},
\label{EV3}
\end{equation}
where $\Gamma $ is the gamma function. They are displayed in Fig.~\ref{F8}c.

\subsection{Third family}

\begin{figure}[t]
\includegraphics[width=4cm]{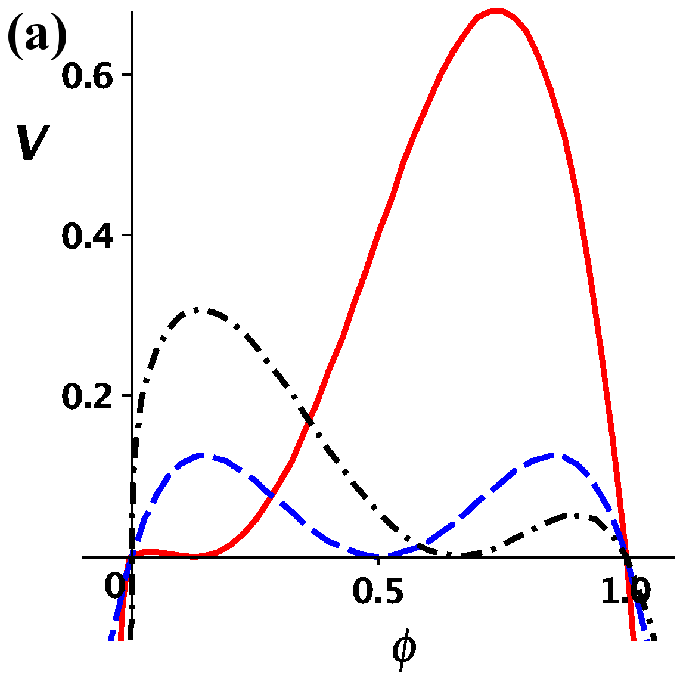}%
\includegraphics[width=4cm]{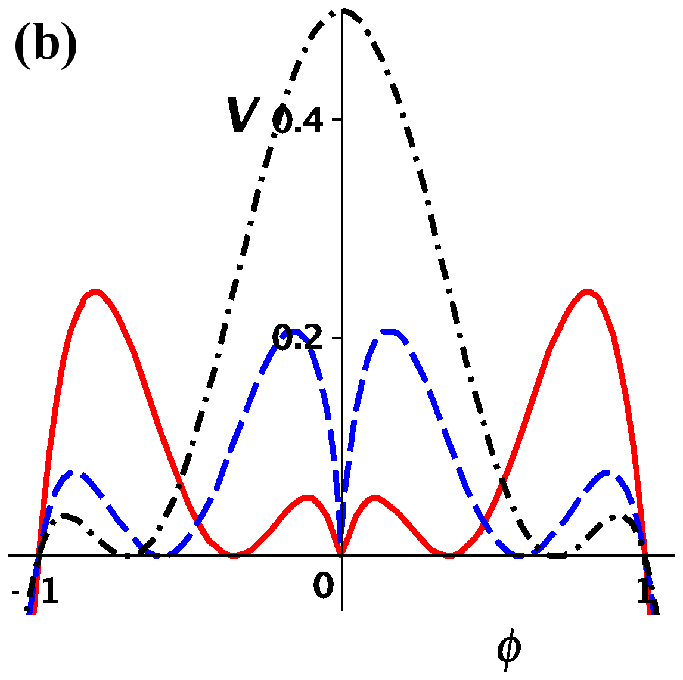}%
\caption{(Color online) Plots of the potential \eqref{V4} for (a) $r=1/3$ with solid (red) line, $r=1$ with dashed (blue) line, and $r=5/3$ with dot-dashed (black) line and for (b) $r=2/3$ with solid (red) line, $r=4/3$ with dashed (blue) line, and $r=2$ with dot-dashed (black) line.}\label{F9}
\end{figure}

Let us now introduce a new family of models, with the potential
\be
V_{4}(\phi )=\frac{2}{r^2}\phi ^{2-r}\left( 1-\phi^{r}\right)(1-2\phi^r)^2 ,
\label{V4}
\ee
where we take $0<r\leq2$, with $r\neq1/m$ for $m$ even. Some local maxima are given by $2^{-1/r}$ and by
\be
\frac{3+r\pm\sqrt{1+2r+5r^2}}{4(1+r)}.
\ee
In Fig.~\ref{F9} we display the potential \eqref{V4} for some specific values of $r$.

We search for solutions following the lines of Sec.~~II. The family of models described by \eqref{V4} has lump-like solutions and energy densities given by
\be
\phi(x)=\frac1{(1+\tanh^2(x))^{1/r}},
\label{solV4}
\ee
and
\be
\rho(x)=\frac{4}{r^2}\frac{\cosh(x)^{\frac{2(2-r)}{r}}\sinh^2(x)}{(2\cosh^2(x)-1)^{\frac{2r+2}{r}}}.
\label{rhoV4}
\ee
These nontopological structures are different from the former ones, since they do not vanish asymptotically. This behavior in nontopological structures also appears in other branches of physics, since the solution now has a pedestal: in magnetically confined plasmas there sometimes appear similar solutions;
see, e.g., \cite{P}. In optics, the technique of nonlinear compression of optical pulses usually leads to a sharp and narrow spike centered on top of a broad low intensity pedestal which carries a large portion of the pulse energy \cite{O}. The output is then similar to the solution found above, with a narrow pulse on top of the pedestal, with the height of the pedestal being nicely controlled by the parameter $r$, which can be of good help for practical use.

\begin{figure}[t]
\includegraphics[width=4cm]{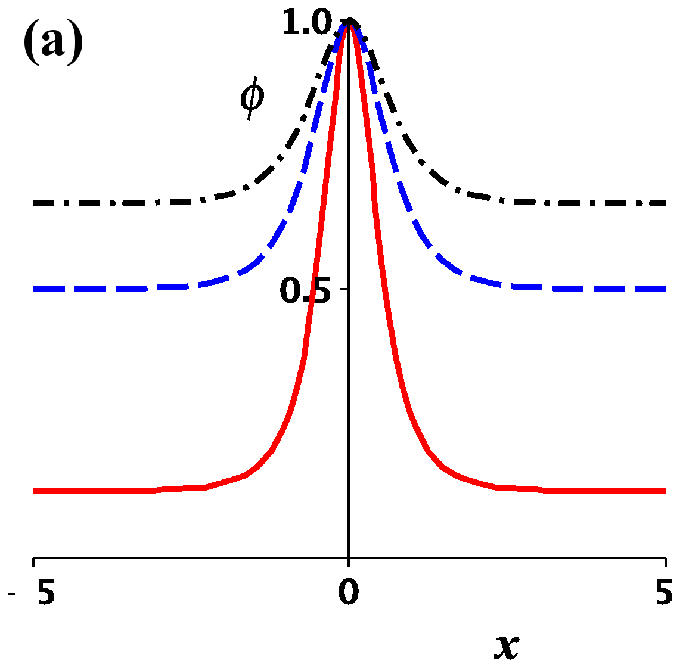}
\includegraphics[width=4cm]{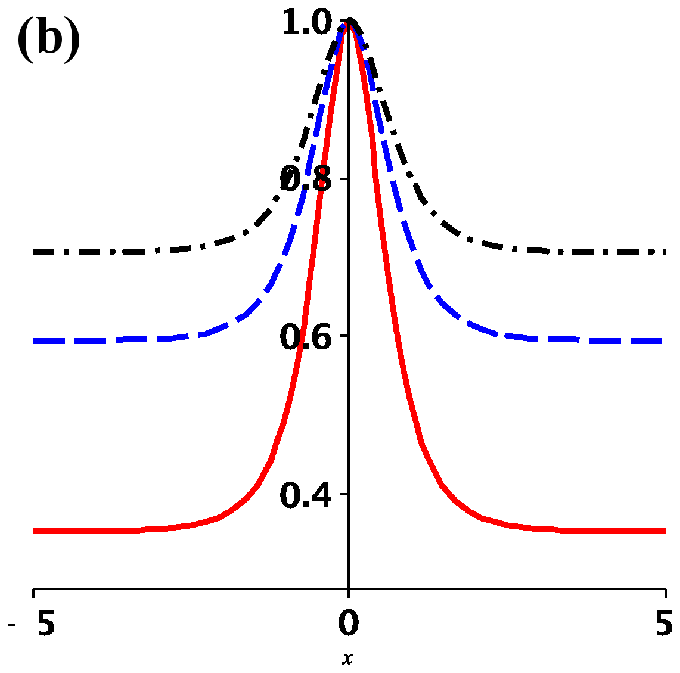}
\caption{(Color online) Plots of the lump-like solutions given by Eq.~\eqref{solV4} for (a) and (b) with the same conventions of Fig.~9.}\label{F10}
\end{figure}

\begin{figure}[t]
\includegraphics[width=4cm]{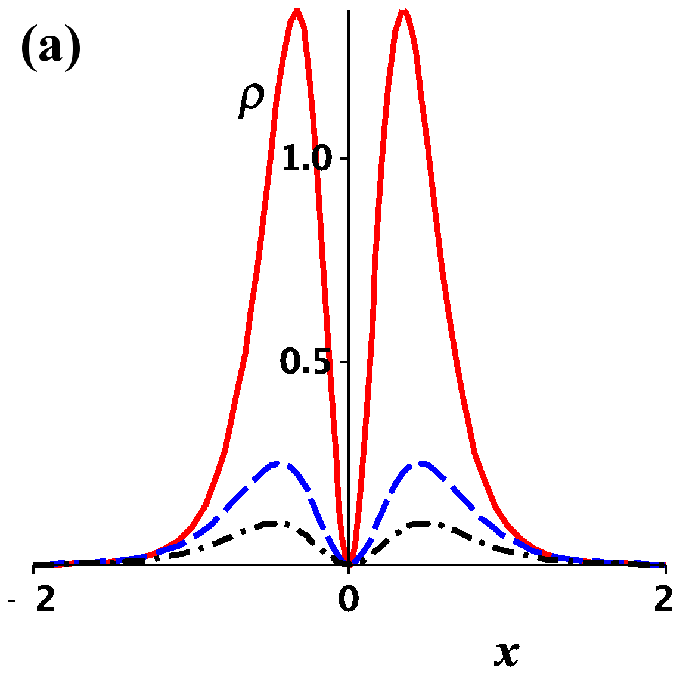}
\includegraphics[width=4cm]{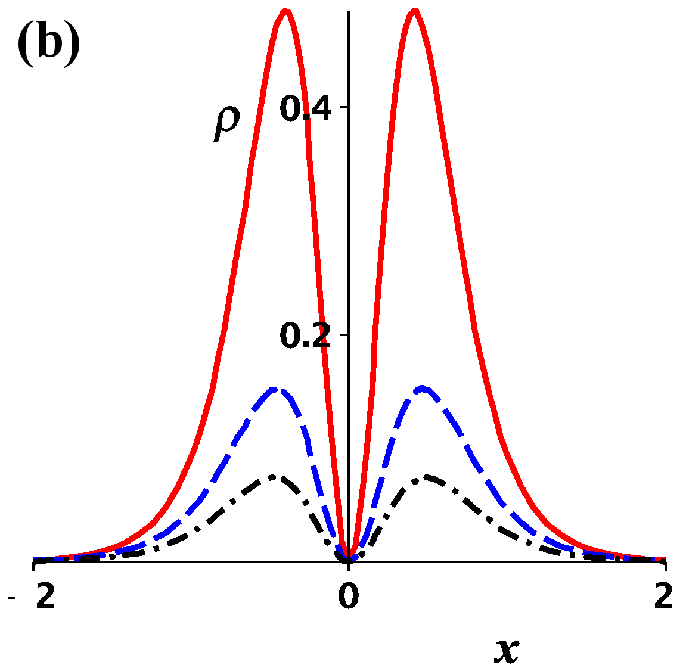}
\newline
\includegraphics[width=4cm]{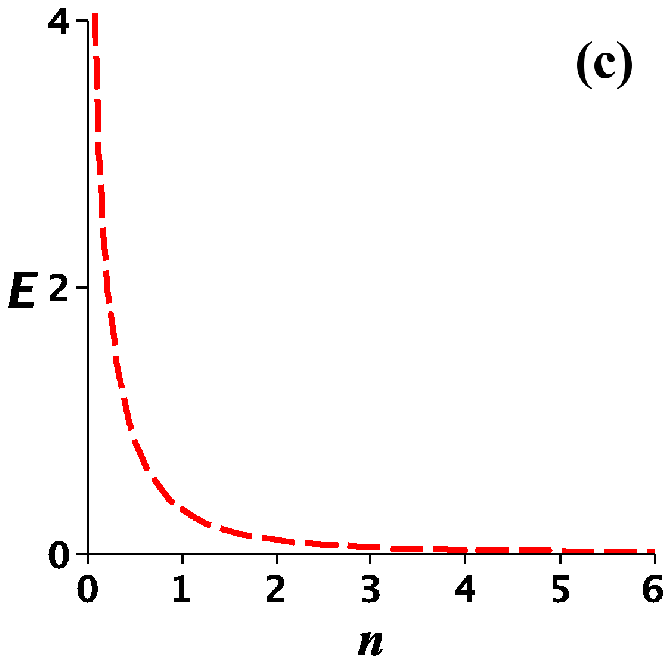}
\caption{(Color online) Plots of the energy density given by Eq.~\eqref{rhoV4} for (a) and (b) with the same conventions of Fig.~9. In (c) we display the total energy computed using \eqref{W00V4} versus $r$. These plots are depicted after removing the pedestal contribution.}\label{F11}
\end{figure}

The amplitude and width of the non-topological bell-shaped solution are $A_{r}=1$ and
\be
L_{r}={\rm arctanh}(\sqrt{2^r-1}),
\ee
without taking into account the pedestal. The solutions for some values of $r$ are presented in Fig.~\ref{F10}a and \ref{F10}b, respectively. The Figs.~\ref{F11}a and \ref{F11}b show the corresponding energy densities, calculated removing the pedestal contribution.

To calculate the total energy we make use of Eq.~(\ref{EW}) to get
\be
W(0)=\frac{2\sqrt{\pi}(r-1)\Gamma\left(\frac{4-r}{2r}\right)}{r^3\Gamma\left(\frac{2(r+1)}{r}\right)},
\label{W0V4}
\ee
and
\ben
W(\infty)&=&\frac{2^{\frac{7r-4}{2r}}}{16-r^2}\nonumber
\\
&&{\rm hypergeom}\left(\!\left[-\frac{1}{2},\frac{4\!-\!r}{2r}\right]\!;\!\left[\frac{3r\!+\!4}{2r}\right]\!;\!\frac{1}{2}\!\right)\!.\;
\label{W00V4}
\een
The total energy of the system is plotted in Fig.~\ref{F11}c, after removing the pedestal contribution.

\section{Ending comments}

In this work we have investigated the presence of bell-shaped lump-like structures in relativistic models described by a single real scalar field in $(1,1)$ spacetime dimensions. These lump-like solutions solve nonlinear equations of motion under the action of boundary conditions which make them nontopological, with bell-shaped profile. They are usually unstable under small fluctuations, but they are of interest in several different contexts \cite{AB,QB1,FB,G,BD,BS,ABC}, including the possibility to generate q-balls, which appear when the discrete symmetry is enlarged to continuum, global symmetry \cite{QB1}. In this case, the potential has to allow for the Noether charge of the elementary excitations present in the model to be greater than the change engendered by the classical solutions.  

Other applications include the use of lump-like excitations in the braneworld context, where they may lead to tachyons living on the brane \cite{FB}, and to gravitating nontopological structures \cite{G}. Recent interest has also been given to issues of stability and fermion modes living on such nontopological structures \cite{BD}. Moreover, we point out the practical use of lump-like models to describe the presence of localized excitations in Bose-Einstein condensates \cite{ABC}.

In this work we have studied several specific models, searching for the lump-like solutions together with the corresponding energies, amplitudes and widths which characterize these solutions. The present study adds new results to the somehow hard subject of proposing new models and finding explicit analytical solutions of nontopological nature. 

In the former work \cite{AB}, we have dealt with other distinct families of models, and the present results show that the procedure of obtaining a first-order framework to study nontopological excitations in relativistic models described by a single real scalar field works very appropriately. As we have shown here and in \cite{AB}, we could explore different families of models, with the nonlinearities playing different roles and making the systems hard to solve, but we have been able to extract the exact solutions for all the models with the explicit help of the first-order framework proposed in \cite{AB}. The results motivate us to further explore the subject, trying to extend the procedure to the case of two or more real scalar fields, and to the case of a complex field. These and other issues are now under consideration, and we hope to report on them elsewhere.

\textbf{\emph{Acknowledgements}} - We would like to thank CAPES, CNPq, FUNAPE-GO, and PRONEX-CNPq-FAPESQ for partial financial support.


\end{document}